\documentclass[final,5p,times,twocolumn]{elsarticle}

\usepackage{amsmath,amsfonts,amssymb}
\usepackage{graphicx} 
\graphicspath{{./images/}}
\usepackage{subcaption}  
\usepackage{epstopdf}
\usepackage{bm}
\usepackage{lineno}
\usepackage{multirow,booktabs}
\usepackage[table,xcdraw]{xcolor}

\biboptions{numbers,sort&compress}
\usepackage{xr-hyper}
\usepackage{hyperref}
\hypersetup{
    colorlinks=true,
    linkcolor=Maroon,
    citecolor=Maroon,
    urlcolor=Maroon,
    breaklinks=true,
}
\usepackage{textcomp,gensymb}
\usepackage{float}
\usepackage{listings}
\usepackage{color}
\definecolor{dkgreen}{rgb}{0,0.6,0}
\definecolor{gray}{rgb}{0.5,0.5,0.5}
\definecolor{mauve}{rgb}{0.58,0,0.82}
\usepackage{placeins} 

\journal{Acta Materialia}

\begin{document}

\begin{frontmatter}

\title{Physics-Informed Gaussian Process Classification for Constraint-Aware Alloy Design}

\author[1]{Christofer Hardcastle}
\author[1]{Ryan O'Mullan}
\author[1,2,3]{Raymundo Arr\'oyave}
\author[1]{Brent Vela \corref{cor1}}

\address[1]{Department of Materials Science and Engineering, Texas A\&M University, College Station, TX 77843, USA}
\address[2]{J. Mike Walker '66 Department of Mechanical Engineering, Texas A\&M University, College Station, TX 77843, USA}
\address[3]{Wm Michael Barnes '64 Department of Industrial and Systems Engineering, Texas A\&M University, College Station, TX 77843, USA}

\cortext[cor1]{Corresponding author email: \texttt{brentvela@tamu.edu}}

\begin{abstract}
Alloy design can be framed as a constraint-satisfaction problem. Building on previous methodologies, we propose equipping Gaussian Process Classifiers (GPCs) with physics-informed prior mean functions to model the boundaries of feasible design spaces. Through three case studies, we highlight the utility of informative priors for handling constraints on continuous and categorical properties. (1) \textbf{Phase Stability:} By incorporating CALPHAD predictions as priors for solid-solution phase stability, we enhance model validation using a publicly available XRD dataset. (2) \textbf{Phase Stability Prediction Refinement:} We demonstrate an \emph{in silico} active learning approach to efficiently correct phase diagrams. (3) \textbf{Continuous Property Thresholds:} By embedding priors into continuous property models, we accelerate the discovery of alloys meeting specific property thresholds via active learning. In each case, integrating physics-based insights into the classification framework substantially improved model performance, demonstrating an efficient strategy for constraint-aware alloy design.
\end{abstract}

\begin{keyword}
Constraint Satisfaction-based Alloy Design \sep Gaussian Process Classification \sep Bayesian Optimization \sep Informed Priors \sep Active Learning
\end{keyword}

\end{frontmatter}

\section{Introduction}

Due to the multitude of performance requirements in materials development, alloy design is often more accurately framed as a constraint satisfaction problem rather than a pure optimization problem \cite{BROUCEK2024120651,yang2024accelerated,xu2024discovering,acemi2024multi}. In this framework \cite{arroyave2016inverse}, the objective shifts from optimizing a single function to identifying one—or all—solutions that satisfy all imposed constraints. This perspective is particularly relevant to alloy design, where the violation of even a single constraint can render a material unsuitable for a specific application. Consequently, it is imperative to develop methods that efficiently navigate feasible design spaces while reducing the reliance on costly experiments \cite{ABUODEH201841,Khatamsaz2023BayesianOW}.
For example, phase stability constraints are particularly common in alloy design \cite{Pollock_Van_2019,peters2024materials}, as specific phases are often desired while deleterious phases need to be avoided. X-ray diffraction (XRD) or microscopy at multiple resolutions are typically employed to determine the presence of various phases in bulk alloy samples. Likewise, high-temperature compression/tension measurements are common objectives in alloy design schemes \cite{qi2024integrated,VELA2023119351,VELA2023118784}  yet are difficult and expensive to execute \cite{skrotzki2018high}.

Due to the combinatorial vastness of alloy design spaces, the time and financial costs of brute-force experimental exploration become prohibitive \cite{Khatamsaz2023BayesianOW}. To alleviate this burden, computational techniques—such as the modified Hume-Rothery rules and CALPHAD-based approaches—have been widely employed as a first approximation for phase stability assessments and predictions \cite{troparevsky2015beyond,li2020high}. Although heuristics like the modified Hume-Rothery rules enable rapid screening of potential single-phase solid solutions, their accuracy is limited for complex multi-component systems; moreover, they cannot predict phase stability as a function of temperature or identify specific intermetallic phases \cite{troparevsky2015beyond}. In contrast, CALPHAD techniques offer higher accuracy but rely heavily on thermodynamic databases \cite{li2020high} that are often labor-intensive to calibrate and less adaptable to the dynamic incorporation of new data in iterative experimental campaigns \cite{zhang2022calphad}. Similarly, in the context of yield strength, several inexpensive analytical models \cite{rao2021theory,maresca2020mechanistic,rao2019solution} predict various strengthening mechanisms; however, these models exhibit limited accuracy when compared to ground-truth experimental measurements.

Recent advances in machine learning have demonstrated significant promise in addressing these challenges. In particular, adaptive models that utilize active learning can dynamically update predictions of material properties as new experimental data become available \cite{dai2020efficient,koizumi2024performance}. Nonetheless, purely data-driven approaches often overlook valuable physical insights, thereby limiting their reliability when data are sparse or incomplete. When alloy design problems are highly constrained, we believe it is more appropriate to frame the design process as a constraint satisfaction problem rather than a pure optimization problem. In our previous work \cite{VELA2023119351,MORCOS2024104545}, we demonstrated that incorporating physics-informed priors into Gaussian Process Regressors (GPRs) significantly improved both the physical accuracy and predictive performance of the models, leading to more efficient Bayesian optimization strategies \cite{VELA2023119351}. In other research \cite{Khatamsaz2023BayesianOW}, we explored how active learning could be used to refine the feasible design space in Bayesian optimization; however, the Gaussian Process Classifiers (GPCs) employed were purely data-driven and lacked informative priors mean functions.

In this study, we address the challenge of dynamically updating predictive models for constrained properties—as new experimental data become available—by proposing a Bayesian classification approach that seamlessly integrates prior knowledge derived from physics-based models. Specifically, we introduce a physics-informed classification method to handle both continuous and categorical constraints in alloy design, targeting properties such as phase stability and yield strength. This approach not only refines predictions with incoming data but also enhances model interpretability and reliability in scenarios where data acquisition is expensive or time-consuming. Moreover, the probabilistic framework enables rigorous quantification of classification uncertainty, which is crucial for informed design and decision-making.

We validate our method through three case studies:
\begin{enumerate}
    \item To demonstrate its utility for categorical classification, we benchmark the proposed method using a publicly available dataset on phase stability in high entropy alloys \cite{MACHAKA2021107346}.
    \item We extend the method to active learning for categorical constraints, demonstrating its ability to construct accurate phase stability predictions with minimal ground-truth data.
    \item Finally, we apply the method to active learning for continuous constraints, specifically yield strength. In this case, equipping Gaussian Process Classifiers (GPCs) with informative priors significantly enhances both classification performance and the active learning of feasible design spaces compared to purely data-driven techniques.
\end{enumerate}

\section{Methods}
\subsection{Gaussian Process Classification for Categorical Data}
In our previous work \cite{VELA2023119351,MORCOS2024104545}, we demonstrated how any Gaussian Process Regressor (GPR) can be equipped with a non-zero prior mean function. This is achieved by training a GPR on the differences between the training data and the prior predictions for that data. Specifically, the model is trained to predict the error in the prior prediction for each data point in the training set. For new data points, the model's predicted error is added to the prior prediction to obtain a final prediction. Mathematically, this approach is equivalent to using a GPR with a non-zero mean function \cite{Bagnell2009GPR}. We found that, on average, models utilizing this method converge during Bayesian optimization faster than standard GPRs trained on the same data \cite{VELA2023119351}. This method can also be extended to classification by adjusting the prior mean function of the latent Gaussian Process (GP) required during GP classification.

All Gaussian Process Classifiers (GPCs) rely on the creation of a latent GP \cite{rasmussen2010gaussian}. This latent function serves as a hidden or nuisance function that describes the underlying process generating the observed class labels. When the latent function is transformed through a sigmoid function it is constrained between 0 and 1 and thus represents the predicted class probabilities. The latent GP captures the knowledge gained from observing the labeled data and acts as a nuisance function—one that is not directly observable but is influenced by the training data through the likelihood function. In the vanilla GPC framework, the label data (encoded as -1 or 1) are related to the latent function via the likelihood function, which is typically modeled using logistic or probit functions. For further details on traditional GPCs, refer to Ref. \cite{rasmussen2010gaussian}. 

Although the rigorous statistical approach to forming GPC with informative prior mean functions is technically sound, the latent function is never directly observed. As a result, its prior mean function cannot be adjusted in a meaningful way. To create GPCs with informative priors, we adopt a less rigorous but practical approach: 

First, we create a latent Gaussian Process (GP). Instead of training it with label data and binary likelihood functions (such as logistic or probit), we train the latent GP using a Gaussian likelihood, as in the case of regression. Specifically, we use label data that spans a wider range, i.e. between \(-5\) and \(5\), rather than being constrained between \(-1\) and \(1\). The GP regressor is trained such that it passes through the training points, where a value of \(5\) represents the positive class, and a value of \(-5\) represents the negative class. 

Once the latent GP regressor is established, it is converted into a GP classifier by passing its outputs through a sigmoid function. This transformation effectively bounds the outputs between \(0\) and \(1\), as shown in Equation~\ref{eq:sigmoid}, where \(f(x)\) is the output of the latent GP, \(\sigma\) is the sigmoid function, and \(p(y=1 \mid x)\) is the probability that \(y\) belongs to class 1 given data \(x\).

\begin{equation}\label{eq:sigmoid}
p(y = 1 \mid x) = \sigma(f(x)) = \frac{1}{1 + e^{-f(x)}}
\end{equation}

\begin{figure*}[htb!]
  \centering
  \includegraphics[width=1\textwidth]{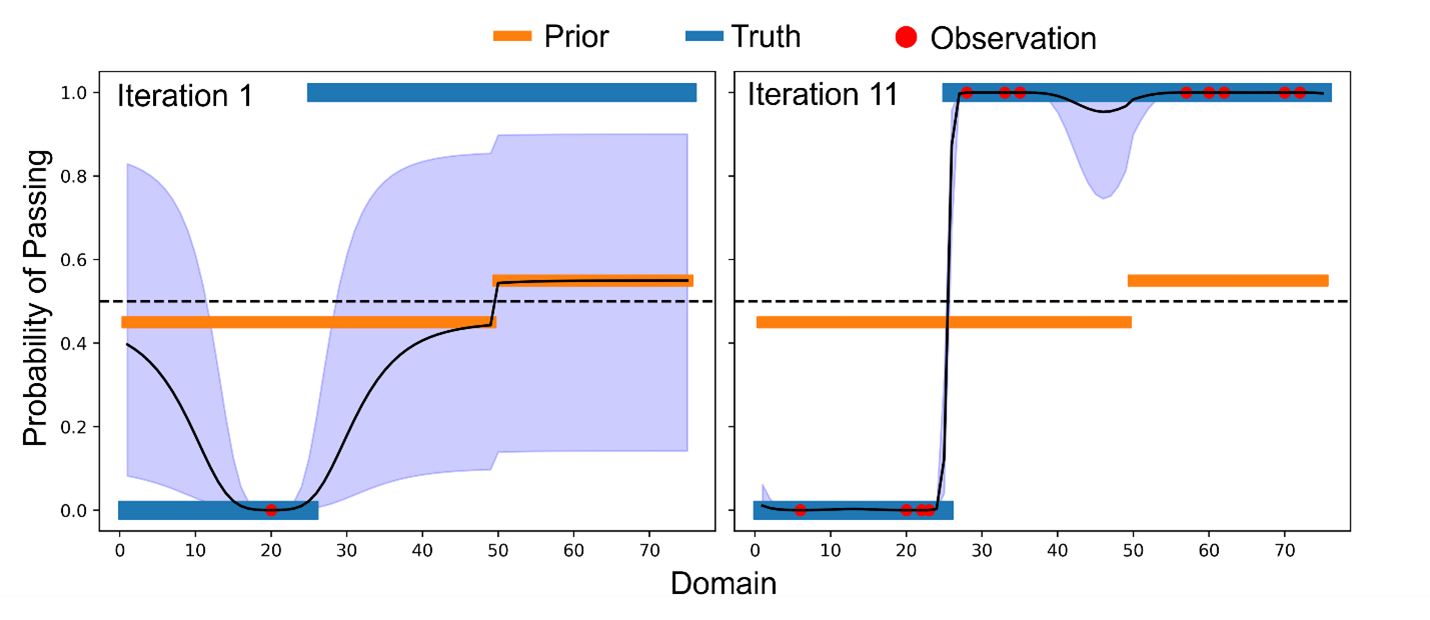}
  \caption{A 1D demonstration of a GPC with an informative prior. The informative prior has a decision boundary at x = 50 while the true decision boundary is at x = 26. In the first iteration a single data point on the left side bolsters confidence in the prior, decreasing the probability of passing to 0 where we have observed a failure. By the 11th iteration the true decision boundary is approximated. There is only a region of uncertain predictions from the classification at $~40 < x < ~50$.}\label{fig:1dtoy}
\end{figure*}

To add an informative prior probability prediction to the GPC, we include an informative prior mean function in the latent GP regressor. For instance, if we want to specify that the prior mean probability is \(55\%\) for Class 1, as is the case in Figure \ref{fig:1dtoy}, the prior mean function in latent space can be defined as \text{logit}(0.55). This logit value serves as the prior mean function for the latent GP. When this value is passed through the sigmoid function, it yields the specified prior probability, as shown in Figure \ref{eq:logit}.

\begin{equation}\label{eq:logit}
\sigma(\text{logit}(0.55)) = 0.55
\end{equation}

This approach effectively incorporates informative priors within a flexible and interpretable Gaussian Process framework. By treating latent GP outputs as continuous values and transforming them into class probabilities, it bridges the gap between regression and classification. The use of GP regressors for classification has precedent; for instance, Dai et al. \cite{dai2020efficient} constructed a GPC using a similar methodology. However, their work did not modify the prior mean function of the latent GP. In contrast, we enhance the prior mean probability prediction by integrating physics-informed models.

To handle multi-class classification, we employ an ensemble of \textit{one-vs-rest} classifiers. In this approach, each class $i$ is associated with its own GPR, which is responsible for predicting the error in the prior probability for that specific class. For each class, the latent GPR predicts the error, which is then added back to the prior prediction. Afterward, the updated prediction is transformed using the sigmoid function to ensure it is bounded between \(0\) and \(1\), yielding valid class probabilities. This process is shown mathematically in Eqn. \ref{eq:p_is_sigma_f}, where \(f_i(x)\) denotes the output from the GPR for class \(i\).

\begin{equation}\label{eq:p_is_sigma_f}
p_i = p(y = i | x) = \sigma(f_i(x))
\end{equation}

Once we have the individual probabilities for each class, $p_i$, we can apply various normalization techniques to generate a multi-class probabilistic prediction for a particular class, $k$. The probabilistic prediction that a data point $y$ belongs to class $k$, $p(y = k | x)$, is the output of an ensemble of one-vs-rest classifiers. This process involves taking the raw output probabilities from each classifier and normalizing them so that they sum to 1, thus transforming the predictions into a valid probability distribution over all classes. This normalization is essential for interpreting the predictions as a set of probabilities. The formula for normalizing the probabilities is shown in Equation \ref{eq:ensemble_normal} where $\sigma(f_k(x))$ is the raw probability (score) from the binary classifier for class \( i \) and the denominator is the sum of all the raw probabilities for all \( n \) classes

\begin{equation}\label{eq:ensemble_normal}
p(y = k | x) = \frac{\sigma(f_k(x))}{\sum_{i=1}^{n} \sigma(f_i(x))}
\end{equation}

This method ensures that the probabilities are properly scaled, but it does not always account for the relative confidence of the classifiers. An alternative approach is to use \emph{softmax} normalization, which normalizes the probabilities and considers each classifier's relative confidence. The \emph{softmax} function converts the raw class probabilities into a distribution where the sum of all probabilities equals 1. This ensures that the resulting probabilities represent the likelihood of each class, making them directly comparable. Additionally, the \emph{softmax} function amplifies the differences between class scores, making it particularly useful when there is a large disparity in classifier confidence. 

The \emph{softmax} function in Eqn. \ref{eq:ensemble_softmax}, where \( p_k \) is the raw probability (logit) from the classifier for class \( k \). The denominator is the sum of the exponentiated probabilities for all \( k \) classes, ensuring that the probabilities sum to 1.


\begin{equation}\label{eq:ensemble_softmax}
p(y = k | x) = \frac{e^{\sigma(f_k(x))}}{\sum_{i=1}^{n} e^{\sigma(f_i(x))}}
\end{equation}




% ...


% ...


% ...




\subsection{Gaussian Process Classification for Continuous Data}\label{sec:cont_data}
Classification can be extended to continuous properties by assessing whether a property exceeds or falls below a specific threshold, such as meeting or failing to meet a property constraint. This approach is particularly relevant in alloy design, where the objective is often to create an alloy that satisfies multiple constraints rather than optimizing a single property \cite{ABUODEH201841}. In these scenarios, it is crucial not only to classify whether the constraints are met but also to quantify the confidence in each prediction. This classification task can be achieved using a GPR.

\begin{figure*}
    \centering
    \includegraphics[width=0.95\linewidth]{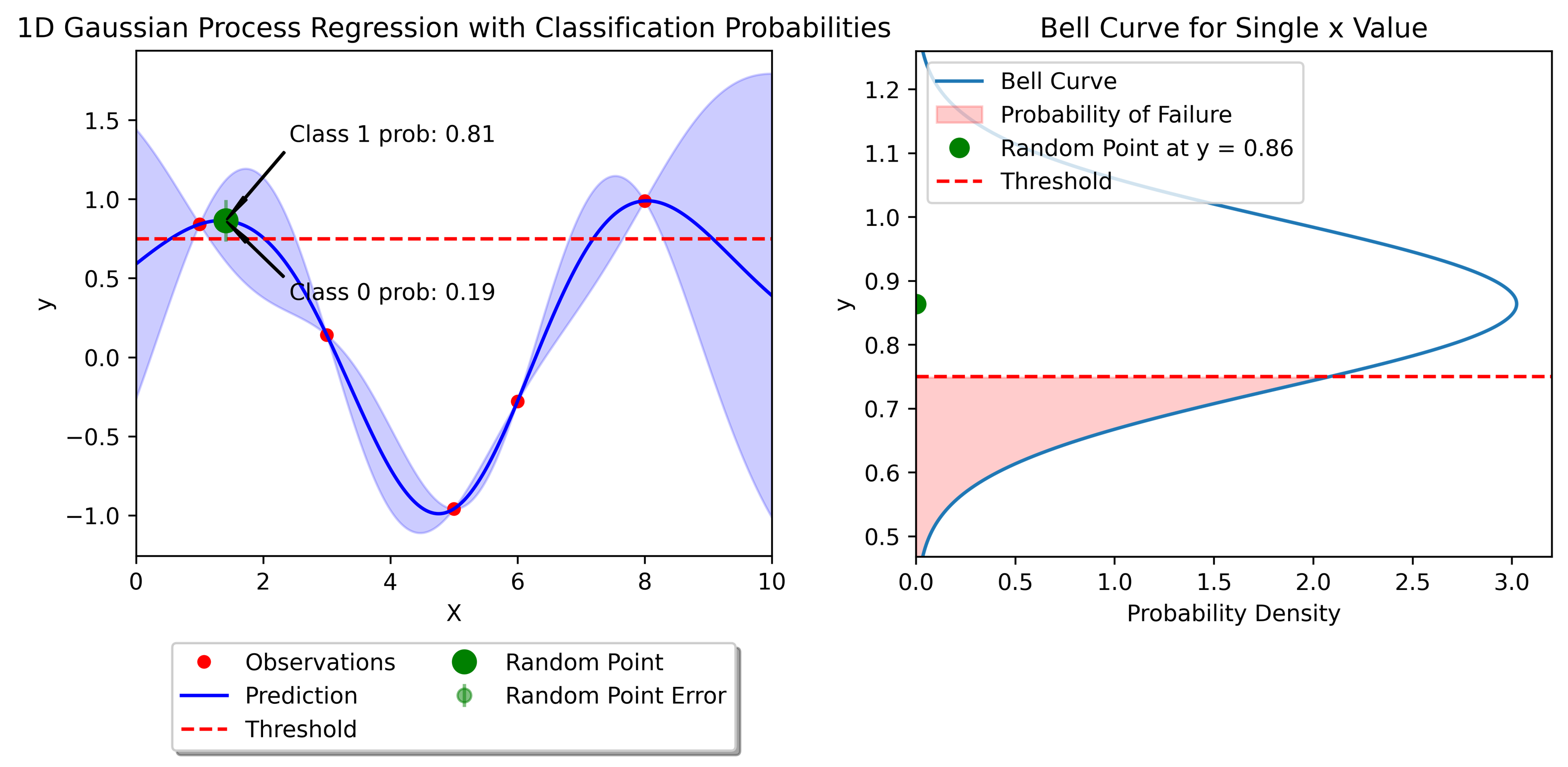}
    \caption{Classification of continuous properties using Gaussian Process Regression (GPR). (a) Illustration of the GPR-based classification process, where red dots represent the limited training observations used to fit the GPR model. The GPR predicts normal distributions for each value of \(x\), and probabilities of meeting or failing a specified threshold are determined using the Cumulative Distribution Function (CDF). Classification is based on whether the probability of meeting the constraint exceeds 0.5, with results visualized across the domain. (b) Visualization of the corresponding bell curve for a single GPR prediction, highlighting the mean prediction and the probabilities of exceeding or falling below the threshold.}\label{fig:cont_class_demo}
\end{figure*}

Consider the example of the classification of continuous properties in Figure \ref{fig:cont_class_demo}. A GPR is trained on a limited number of observations (red dots). Based on these observations, the GPR will interpolate and extrapolate $y$ values across the $x$ domain. Predictions from GPRs are normal distributions. For each value of $x$ in the domain, the GPR returns the mean prediction and standard deviation (each prediction is Gaussian and is determined by the posterior distribution over functions \cite{rasmussen2010gaussian}). Since each prediction is a normal distribution, the probability that a property falls below a threshold can be found using the Cumulative Distribution Function (CDF). Similarly, the probability that a property exceeds a threshold can be found by subtracting the CDF from 1. Once the probability of exceeding or falling below a threshold is determined, the property is classified as meeting the constraint if the probability is greater than 0.5. Otherwise, it is classified as failing to meet the constraint.

\subsection{Classification Error Metrics}\label{sec:error_metrics}

To evaluate model performance, we calculated six classification metrics for predictions on the test subsets: accuracy, precision, recall, F1-score, log-loss, and multi-class Brier loss. Accuracy measures the proportion of correctly classified samples, as defined in Eqn. \ref{eq:accuracy} where \( TP \), \( TN \), \( FP \), and \( FN \) represent the counts of true positives, true negatives, false positives, and false negatives, respectively. Precision quantifies the fraction of predicted positive cases that are true positives, as defined in Eqn. \ref{eq:precision}. Recall indicates the proportion of correctly identified positive cases, as defined in Eqn. \ref{eq:recall}. The F1-score combines precision and recall into a single harmonic mean to summarize the test's accuracy, providing a balanced measure that accounts for both false positives and false negatives, as defined in Eqn. \ref{eq:f1}. Log-loss (Eqn. \ref{eq:logloss}) and Brier loss (Eqn. \ref{eq:brierloss}) evaluate the accuracy of predicted probabilities by penalizing incorrect confidence levels. In these equations \( N \) is the total number of data points, \( R \) is the number of classes, \( f_{ti} \) is the predicted probability of class \( i \) for data point \( t \), and \( o_{ti} \) is 1 if \( t \) belongs to class \( i \), otherwise 0.

\begin{equation}\label{eq:accuracy}
\text{Accuracy} = \frac{TP + TN}{TP + TN + FP + FN},
\end{equation}

\begin{equation}\label{eq:precision}
\text{Precision} = \frac{TP}{TP + FP}.
\end{equation}

\begin{equation}\label{eq:recall}
\text{Recall} = \frac{TP}{TP + FN}.
\end{equation}

\begin{equation}\label{eq:f1}
F1 = 2 \cdot \frac{\text{Precision} \cdot \text{Recall}}{\text{Precision} + \text{Recall}}.
\end{equation}

\begin{equation}\label{eq:logloss}
\text{Log-Loss} = -\frac{1}{N} \sum_{t=1}^{N} \sum_{i=1}^{R} o_{ti} \log(f_{ti}),
\end{equation}

\begin{equation}\label{eq:brierloss}
BL = \frac{1}{N} \sum_{t=1}^{N} \sum_{i=1}^{R} (f_{ti} - o_{ti})^2.
\end{equation}

\section{Case Study: Benchmark Against Experimental Data}

First, to demonstrate the benefit of informative priors in static classification examples \cite{MACHAKA2021107346}, we benchmark our proposed method against a dataset of experimentally classified phase stability data. Specifically, we utilize a dataset of experimentally labeled phase stability data and their corresponding homogenization temperatures, and employ CALPHAD models to predict the expected equilibrium phases under these conditions. The dataset used in this work is provided in the code repository associated with this work. These CALPHAD predictions are then treated as the prior for probabilistic classification. Next, the database is shuffled and split into training and test sets. Using the training set, the prior probabilities (derived from the CALPHAD phase predictions) for the test set are updated based on the training data (experimental phase labels). The accuracy, precision, recall, F1-score, Brier-loss, and log-loss scores are computed for both the multi-class scenario and several one-vs-rest scenarios. Our method outperforms both the CALPHAD prior model and purely data-driven “vanilla” GPCs.

\subsection{Experimental Dataset}

In this experimental case study, we evaluated the predictive performance of GPCs with informative priors by comparing their predictions to experimental phase stability data. The dataset, curated by Machaka et al. \cite{MACHAKA2021107346}, provides comprehensive information on the phase stability of various High Entropy Alloys (HEAs), including details on alloy synthesis methods, processing conditions (e.g., cold or hot work), heat treatment temperatures, and the resulting phases. To minimize confounding factors, we filtered the dataset to include only as-cast alloys that underwent homogenization heat treatments, excluding those subjected to further processing such as hot or cold working. This filtering was applied to approximate equilibrium conditions, aligning with the predictive capabilities of CALPHAD-based prior models, which focus on equilibrium phase stability. This approach ensures that the experimental data is consistent with the assumptions of the computational framework.

Although the original dataset categorized alloys into seven phase labels, this study focused on the four most common: single-phase FCC alloys, FCC alloys with secondary phases (FCC + Sec.), single-phase BCC alloys, and BCC alloys with secondary phases (BCC + Sec.). Although this proposed method can accommodate classification problems beyond 4 classes, insufficient data for the remaining three labels, particularly after filtering, required this simplification. For the purposes of this study, a 4-label classification framework provides a robust benchmark to validate the proposed method. After filtering, the dataset contained 86 usable data points: In order to facilitate reproducibility and further research, the cleaned and processed dataset is publicly available in the Code Repository associated with this work.

\subsection{Physics-Informed Prior for Phase Stability}

The source of prior information for phase stability was the `Calculation of Phase Diagrams' (CALPHAD) predictions generated using Thermo-Calc’s Python equilibrium module \cite{thermocalc}. This module utilizes the TCHEA6 thermodynamic database \cite{tchea6}, which was specifically chosen for its suitability in modeling compositionally complex alloys. In our previous work \cite{VELA2023119351}, we rigorously evaluated the accuracy of Thermo-Calc’s equilibrium module for predicting phase stability and demonstrated its reliability for alloy design applications. Based on these findings, we considered this module a robust and credible source of prior information for this case study.

Phase stability predictions were generated using Thermo-Calc for each alloy in the filtered dataset at its respective homogenization/heat treatment temperature, representing the equilibrium phases expected under those conditions. Although cooling rates can affect phase formation in practice, these predictions are used solely as prior information and are refined by experimental data. We acknowledge that factors like cooling rates can introduce confounding effects that sometimes reduce the accuracy of Thermo-Calc predictions; however, equilibrium CALPHAD predictions provide a reasonable initial approximation for phase stability—an approximation that can be updated in light of experimental data. In fact, correcting the prior model with data is the main goal of the proposed framework.

The Thermo-Calc equilibrium module predicts the mole fractions of various microstructures. The prior phase classification from Thermo-Calc was assigned according to the following rules:  
\begin{itemize}
    \item If the FCC mole fraction for a data point is $\phi_{\text{FCC}} \geq 0.99$, it is classified as single-phase FCC.  
    \item If $\phi_{\text{FCC}} \geq 0.5$ but less than 0.99, it is classified as FCC with a secondary phase (FCC + Sec.).  
    \item The same thresholds are applied to BCC mole fractions for classification as single-phase BCC or BCC + Sec.  
\end{itemize}

After establishing phase predictions from Thermo-Calc, we quantified our confidence in these prior class predictions using class probabilities. These probabilities reflect the level of certainty associated with a particular classification, whether derived from a vanilla GPC or an informed GPC. In the case of an uninformed GPC, the prior class probability is 50\%/50\%. For an informed GPC, the prior class probability is assigned according to the designer’s judgment. An example of this informed prior class probability is shown in Figure \ref{fig:1dtoy}.

The prior probabilities are detailed in Table \ref{tab:prior_weights}. For instance, if the prior classification for an alloy is single-phase FCC, the confidence is distributed as follows: a 50\% probability of being single-phase FCC, a 40\% probability of being FCC with secondary phases, and a 5\% probability of either being single-phase BCC or BCC with secondary phases. These prior probabilities are intuitive because if Thermo-Calc predicts an alloy to be single-phase FCC, the highest prior probability is assigned to the FCC class. However, because secondary phases may form within the FCC matrix during cooling, the FCC+Sec. class is assigned the second-highest probability. Conversely, if an alloy is predicted to be FCC by Thermo-Calc, it is unlikely to exhibit a BCC matrix experimentally. In other words, while we trust Thermo-Calc’s ability to distinguish between FCC and BCC, we are less confident in its ability to differentiate between FCC and FCC+Sec. and to differentiate between BCC and BCC+Sec.

\begin{table}[t]
    \centering
    \caption{Prior class probabilities based on Thermo-Calc phase predictions.} 
    \label{tab:prior_weights}
    \resizebox{\columnwidth}{!}{ 
        \begin{tabular}{ l l c c c c }
        \toprule
        & & \multicolumn{4}{c}{\textbf{Prior Probability}}  \\ 
        &  & FCC & FCC+Sec.  & BCC & BCC+Sec.    \\ 
        \midrule
        \parbox[t]{2mm}{\multirow{4}{*}{\rotatebox[origin=c]{90}{\textbf{Prior Pred.}}}}&   FCC & \cellcolor{red!70}50\% & \cellcolor{orange!40}40\% & \cellcolor{blue!20}5\% & \cellcolor{blue!20}5\%    \\ 
        &   FCC+Sec. & \cellcolor{orange!40}40\% & \cellcolor{red!70}50\% & \cellcolor{blue!20}5\% & \cellcolor{blue!20}5\%   \\ 
        &  BCC & \cellcolor{blue!20}5\% & \cellcolor{blue!20}5\% & \cellcolor{red!70}50\% & \cellcolor{orange!40}40\%    \\ 
        & BCC+Sec. & \cellcolor{blue!20}5\% & \cellcolor{blue!20}5\% & \cellcolor{orange!40}40\% & \cellcolor{red!70}50\% \\
        \bottomrule
        \end{tabular}
    }
\end{table}

\subsection{Training the Gaussian Processes Classifiers}\label{sec1:Training}

The Gaussian Process Classifiers (GPCs) employed in this work were constructed using the Gaussian Processes (GPs) implementation from Scikit-Learn \cite{scikit-learn}. Although more advanced GP libraries are available, we chose Scikit-Learn for its simplicity and accessibility, which makes it easier for a broader audience to adopt our approach.

To implement the informative GPCs described in the Methods section, we developed a custom class based on Gaussian Process Regressors (GPRs). For each one-vs-rest GPC, a latent GPR is first trained on the class observations, where positive class observations are set to $y = 5$ and negative ones to $y = -5$. This latent GPR is then passed through a sigmoid transformation to constrain the outputs between 0 and 1, yielding valid probabilities. The code for this implementation is available in the associated Code Ocean repository.

The GPRs used in this active learning scheme employ an additive kernel composed of a Radial Basis Function (RBF) kernel and a White Noise (WN) kernel, as defined in Equation~\ref{eq:rbf_noise_kernel}. In Equation~\ref{eq:rbf_noise_kernel}, \( k(\mathbf{x}, \mathbf{x}') \) represents the covariance function between input points \( \mathbf{x} \) and \( \mathbf{x}' \). The first term corresponds to the RBF kernel, where \( \sigma_f^2 \) is the signal variance, controlling the amplitude of function variations, and \( \ell \) is the characteristic length scale, determining how quickly correlations decay with distance. The second term accounts for white noise, where \( \sigma_n^2 \) is the noise variance, and \( \delta(\mathbf{x}, \mathbf{x}') \) is the Kronecker delta function. Selecting an appropriate kernel is inherently challenging and often depends on expert judgment; this choice implicitly assumes specific correlation patterns and functional shapes. The RBF + WN additive kernel is a standard choice that works well in practice.

Kernel hyperparameters were optimized by maximizing the log-marginal likelihood using the L-BFGS-B algorithm as implemented in Scikit-Learn \cite{scikit-learn}. To ensure robust optimization, we performed 10 optimizer restarts for each GPR. For the RBF kernel, the optimization was constrained to search for length scales between 5 atomic percent (at\%) and 100 at\%. This range was chosen based the observation that barycentric spaces cannot have length scales exceeding 100 at\%. These constraints help ensure that the kernel parameters remain physically meaningful and aligned with the characteristics of the sampled data.


\begin{equation}
k(\mathbf{x}, \mathbf{x}') = \sigma_f^2 \exp\left(-\frac{\|\mathbf{x} - \mathbf{x}'\|^2}{2\ell^2}\right) + \sigma_n^2 \delta(\mathbf{x}, \mathbf{x}')
\label{eq:rbf_noise_kernel}
\end{equation}

Table \ref{tab:wen_feats} summarizes the 10 alloy features used to train the model. 
For brevity, specific details on these features can be found at Ref \cite{WEN2019109}. All features are functions of an alloy's chemical composition and were calculated using Matminer’s \texttt{WenAlloys} featurizer \cite{matminer}. These were determined to be useful in predicting solid solution phase stability by Wen et al. \cite{WEN2019109}. While more sophisticated feature selection could be performed, this work aims to highlight the effect of physics-informed prior mean functions during GP classification and does not necessarily identify the most relevant features for phase classification.

\begin{center}\label{tab:wen_feats}
\captionof{table}{Alloy features used to train GPCs.} 
\begin{tabular}{ c | c }
\hline
 Yang delta & Yang omega \\ 
 APE mean & Radii local mismatch \\  
 Radii gamma & Configuration entropy \\
 Atomic weight mean & Total weight \\
 Lambda entropy & Electronegativity delta 
\end{tabular}
\end{center}

\subsection{Experimental Benchmarking Results}\label{sec:machaka_results}

Three models were benchmarked to illustrate the impact of incorporating a physics-informed prior into GPCs: two control models and the proposed model. The first control model was a GPC with an uninformed prior mean function, which assigned equal probabilities (25\%) to all predictions in the four-class case. In the case of four-class classification problems, 25\% probability for all 4 classes represents the state of maximum information entropy, reflecting the highest level of uncertainty in predictions. The second control model consisted solely of the CALPHAD phase predictions. The third model was a GPC with a prior mean function defined by CALPHAD phase predictions. The values of these priors are reported in Table \ref{tab:prior_weights}. As mentioned in Section \ref{sec1:Training} the latent GPs in both GP classifiers were equipped with the same kernel and training settings as the uninformed GPC, specifically the RBF + WN additive kernel, as described in Eqn. \ref{eq:rbf_noise_kernel}. 

Benchmarking the models on a small dataset necessitated the use of cross-validation. We employed stratified Monte Carlo cross-validation, generating 500 random 20\%/80\% train/test splits. This approach differs from the more typical 80\%/20\% splitting and reflects the reality of data-sparse scenarios in alloy design, where experimental data is often prohibitively expensive to collect. Stratification was crucial to maintain the class ratio in both training and testing subsets, ensuring consistency across splits.

\begin{figure*}
    \centering
    \includegraphics[width=0.85\linewidth]{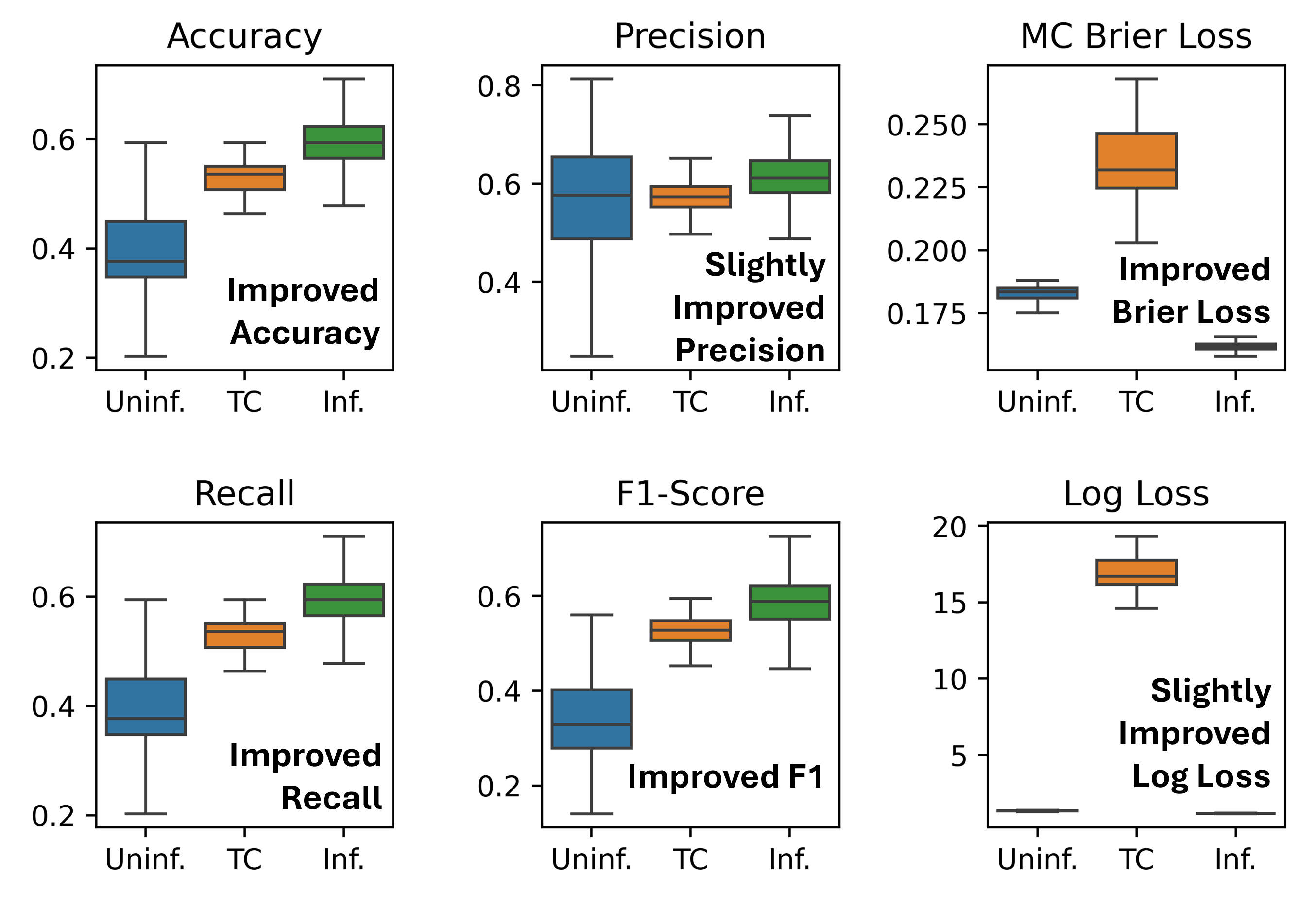}
    \caption{Model errors for the standard GPC (Uninf.), Thermo-calc (TC), and the GPC with the physics-informed prior (Inf.) when predicting across all phases.}
    \label{fig:error_Metrics_ALL}
\end{figure*}

Using box-and-whisker plots to display each error metric across the cross-validation splits, Figure \ref{fig:error_Metrics_ALL} summarizes the overall predictive performance of the three models across all classes. In the context of predicting phase stability as a 4-class classification problem, it is evident that the informed model exhibits, on average, improved accuracy and recall. Although the median precision values of the uninformed and informed classifiers are similar, the interquartile range (IQR) indicates that the CALPHAD-informed model performs more consistently, whereas the uninformed model displays greater variability—an undesirable outcome. We prefer that models perform well and perform well consistently. Furthermore, employing any GPC is preferable to using a model with unquantified uncertainty in its predictions (i.e., a non-probabilistic model).

As clearly demonstrated by the plots, the GPC with the physics-informed prior outperforms both control models on most metrics. The interquartile ranges for accuracy, recall, F1-score, and Brier loss show significant improvements over the control models, with more subtle enhancements in precision and log-loss. To further evaluate each model's ability to correctly identify specific classes, separate analyses of the predictions over the 500 splits were performed and are reported in the Supplemental Material.

\section{Case Study: Active Learning of Categorical Constraints}

Phase stability constraints are particularly common in alloy design, where specific phases are desired and deleterious phases need to be avoided \cite{vela2022evaluating,VELA2023118784,pollock2019evolving}. To assess phase stability, X-ray diffraction (XRD) experiments are typically employed to determine the presence of various phases in bulk alloy samples. However, given the large number of candidate alloys, even ternary alloy systems, the time and cost associated with XRD experiments make a brute-force experimental exploration infeasible \cite{Khatamsaz2023BayesianOW}. To mitigate this burden, computational techniques have emerged to complement experimental efforts in alloy design.

Simple heuristics, such as modified Hume-Rothery rules, have been extended to screen for alloys, particularly medium and high entropy alloys, that form single-phase solid solutions \cite{troparevsky2015beyond}. These methods are computationally inexpensive, allowing for rapid preliminary screening of large compositional spaces \cite{li2020high,wu2021design}. However, these heuristics for phase stability have shortcomings. Their accuracy is often limited \cite{troparevsky2015beyond}. Moreover, these heuristics cannot predict phase stability as a function of temperature. Furthermore, these modified Hume-Rothery rules are only valid in determining if HCP, FCC, BCC or intermetallic phases are likely to form, however these metrics do not provide details about what intermetallic phase is likely to form.

Beyond simple heuristic models, CALculation of PHAse Diagram (CALPHAD) techniques have been employed to predict phase stability in HEA design, particularly in high-throughput computational workflows \cite{li2020high}. The accuracy of CALPHAD predictions relies heavily on the quality and relevance of the underlying thermodynamic databases. CALPHAD databases require careful calibration of parameters to match experimental results. This restricts their applicability in closed-loop experimental alloy design campaigns, where data are dynamic and must be quickly incorporated into models to inform subsequent experiments.

Recent advances in machine learning have demonstrated the potential for on-the-fly updating of phase stability models during experimental campaigns. Machine learning models, particularly those used for classification, can be continuously trained as new data become available, allowing for adaptive, data-driven optimization strategies \cite{dai2020efficient,koizumi2024performance,zhu2024active}. This is known as active learning (AL) of constraints. However, these approaches often neglect valuable physical insights and can suffer from a dependence on large amounts of training data, limiting their effectiveness when data are sparse or incomplete. 


Physics-constrained active learning of phase diagrams have been achived using graph-based techniques such as in the CAMEO framework \cite{kusne2020fly}. Of particular interest to this work, Ament et al. \cite{ament2021autonomous} employed a physics-informed kernel within a GP-based active learning framework to accelerate the construction of phase diagrams by incorporating prior physical knowledge into the model's covariance structure. While this approach has its merits, our work introduces a novel and complementary strategy: incorporating physics through the modification of the GP prior mean function. Since a GP is fully defined by both its mean and covariance functions, embedding domain-specific physical insights directly into the prior mean offers an alternative pathway for guiding predictions—especially beneficial when fast-acting prior models for specific properties are available. In contrast to kernel modification, which is better suited for capturing global trends and symmetries \cite{ament2021autonomous}, adjusting the prior mean function provides a more targeted method for integrating known local physical behaviors.


In this \emph{in silico} case study, we address the challenge of dynamically updating phase stability models as new experimental data become available. Here, the valence electron concentration (VEC) serves as the prior belief regarding the stability of FCC and BCC phases in the Fe-Ni-Cr alloy system at 1000°C. The ground truth for phase stability is provided by Thermo-Calc equilibrium calculations, using the TCHEA6 high entropy alloy database \cite{tchea6}. The objective of this case study is to construct the most accurate isopleth phase stability predictions with the fewest possible queries of the ground truth. Our results demonstrate that active learning schemes incorporating simple yet informative priors outperform those relying solely on vanilla GPCs. This approach aligns with recent efforts to develop closed-loop alloy design frameworks, making this a well-motivated case study.

\subsection{Models for Prior and Ground-Truth}
The Valence Electron Concentration (VEC) of an alloy is defined as the weighted sum of the valence electron concentrations of its constituent elements. Numerous studies \cite{basu2023new,arroyave2022phase,alam2023revisiting,guo2011effect} have shown that VEC is an effective descriptor for predicting single-phase stability and for delineating the boundary between FCC and BCC phase stability. In particular, alloys with a VEC above 8 tend to exhibit FCC structures, while those with a VEC below 6.87 are typically BCC \cite{guo2011effect}. This suggests that alloys with VEC values between 6.87 and 8 are likely to display dual-phase (BCC + FCC) behavior. The formula for calculating VEC is given in Eqn. \ref{ref:eqn}, where \( c_i \) represents the atomic fraction of element \( i \) and \( v_i \) is the valence electron concentration of element \( i \).

\begin{equation}\label{ref:eqn}
\text{VEC} = \sum_{i=1}^{n} c_i v_i    
\end{equation}

We assign prior probabilities based on predictions from the prior model (i.e., the VEC). These probabilities are detailed in Table \ref{tab:prior_weights_AL}. For example, if an alloy has a VEC greater than 8, our degree of belief that the alloy is FCC is represented by a 54\% probability. Conversely, we assign a 23\% probability each to the alloy being dual-phase or BCC.

\begin{table}[h!]
\centering
\caption{Prior Weights}
\label{tab:prior_weights_AL}

\newcommand{\colorcell}[1]{
    \ifdim #1 pt = 54 pt \cellcolor{red!60}
    \else \ifdim #1 pt = 23 pt \cellcolor{blue!30}
    \else \cellcolor{blue!10}
    \fi\fi
    #1\%
}

\renewcommand{\arraystretch}{1.5} 

\begin{tabular}{ l l c c c }
\toprule
& & \multicolumn{3}{c}{\textbf{Prior Probability}}  \\ 
& & FCC & Dual & BCC \\ 
\midrule
\multirow{3}{*}{\rotatebox[origin=c]{90}{\textbf{Prior Pred.}}} & 
FCC & \colorcell{54} & \colorcell{23} & \colorcell{23} \\ 
& Dual & \colorcell{23} & \colorcell{54} & \colorcell{23} \\ 
& BCC & \colorcell{23} & \colorcell{23} & \colorcell{54} \\ 
\bottomrule
\end{tabular}

\end{table}

Regarding the ground truth model for this \emph{in-silico} example, we consider Thermo-Calc's equilibrium calculator—equipped with the TCHEA6 database \cite{tchea6}—as the ground truth. We queried this calculator at 1000$^{\circ}$C for all candidate alloys, which yielded the decision boundaries (i.e., phase boundaries) shown as black dashed lines in Figure \ref{fig:active_learning_results}. The code for the ground-truth model is available in the repository associated with this work.

\subsection{Gaussian Processes and Active Learning Parameters}

The Gaussian Process Regressors (GPRs) used in this active learning scheme employed an additive kernel that combines a Radial Basis Function (RBF) kernel with a White Noise kernel (WN), as defined in Eqn~\ref{eq:rbf_noise_kernel}. Selecting an appropriate kernel is inherently challenging and often relies on expert judgment, since this choice implicitly assumes specific correlation patterns and functional shapes for the underlying process. The RBF + WN additive kernel is a standard choice in such applications.

The kernel hyperparameters were optimized by maximizing the log-marginal likelihood using the L-BFGS-B algorithm, as implemented in scikit-learn. To ensure robust optimization, we performed 50 optimizer restarts for each GPR. The first run used the kernel's initial parameter estimates, while the remaining runs initialized parameters by sampling log-uniformly from the allowed parameter space, ensuring thorough exploration.

For the RBF kernel, the optimization was constrained to search for length scales between 5 atomic percent (at\%) and 100 at\%. Again this range was chosen based on the fact that that barycentric spaces do not have properties that vary at lengthscale greater than 100 at\%. These constraints ensured that the kernel parameters remained physically meaningful and aligned with the characteristics of the sampled data. The active learning framework for categorical properties used a GPfor the surrogate model and maximum Shannon entropy for the aquisition function \cite{chen2020active}.

\subsection{Categorical Active Learning  Case Study Results}

\begin{figure*}
    \centering
    \includegraphics[width=0.95\linewidth]{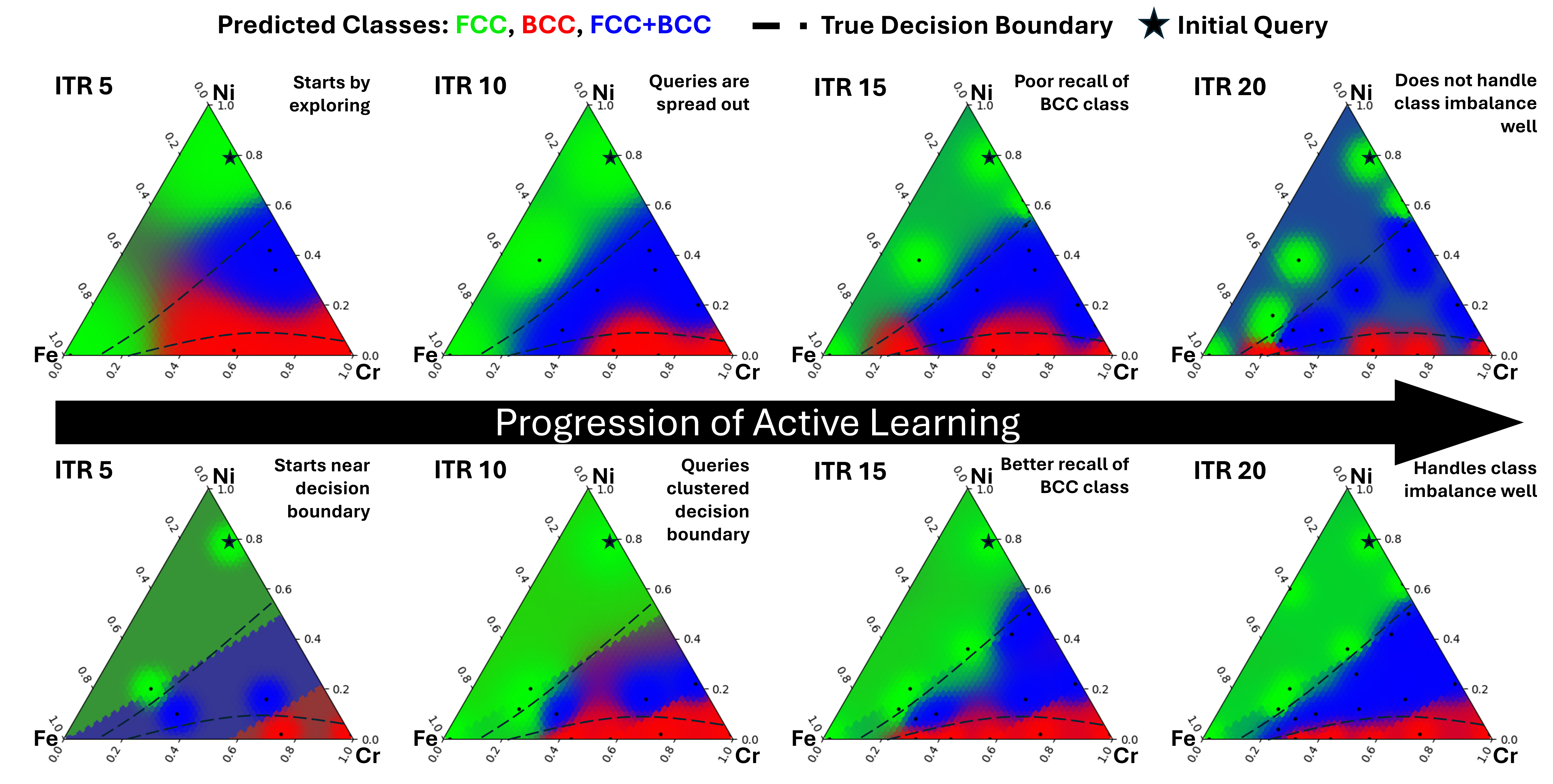}
    \caption{Comparison of vanilla and physics-informed Bayesian active learning for phase stability predictions in the Fe-Ni-Co system at 1000$^{\circ}$C. The top row displays the vanilla AL scheme, while the bottom row shows the physics-informed AL scheme. Colors represent class probabilities via an RGB scheme, with green indicating FCC, blue indicating FCC+BCC, and red indicating BCC. In early iterations, the physics-informed model heavily relies on its prior knowledge. By iteration 15, it significantly improves recall for the BCC phase, and by iteration 20, it demonstrates greater robustness to class imbalance compared to the vanilla approach, achieving more precise decision boundaries.}
    \label{fig:active_learning_results}
\end{figure*}

To evaluate the impact of informative priors on Bayesian active learning for phase stability predictions within the Fe-Ni-Co system at 1000$^{\circ}$C, we compared a physics-informed active learning scheme to a physics-uninformed scheme. Each scheme operated under a fixed budget of 25 queries to the ground truth per active learning campaign. Figure \ref{fig:active_learning_results} shows an example of a single active learning campaign. Class probabilities in this 3-class scenario are visualized using an RGB color scheme. For instance, if an alloy is predicted to be FCC with 100\% confidence (i.e., a probability of 1.0), the corresponding RGB value is [0, 255, 0], resulting in a bright green color in the ternary diagram. Similarly, alloys predicted to be dual-phase with 100\% confidence are plotted as blue (RGB = [0, 0, 255]). If the model predicts equal probabilities for all three phases (33\%/33\%/33\%), the RGB value is [85, 85, 85], and the alloy is displayed as gray, indicating the highest Shannon entropy and, consequently, the greatest uncertainty in the prediction \cite{chen2020active}.

The top row shows the progression of the vanilla active learning (AL) campaign, while the bottom row displays that of the physics-informed AL campaign. At the \textbf{5$^{th}$ iteration}, the physics-informed approach already leverages its prior knowledge (e.g., phase predictions from the VEC) to accurately delineate the decision boundary between the FCC and dual-phase regions, though it still struggles to separate the dual-phase from the BCC region. At the \textbf{10$^{th}$ iteration}, the physics-informed model achieves better recall for the BCC class than the vanilla model; however, predictions in the BCC region are rendered in purple, indicating uncertainty between a pure BCC phase and a mixed FCC+BCC state—while clearly ruling out single-phase FCC (green). By the \textbf{15$^{th}$ iteration}, the physics-informed scheme further refines its predictions, markedly improving recall for the minority BCC class. Finally, at the \textbf{20$^{th}$ iteration}, the vanilla AL scheme reveals its limitations in handling class imbalance by heavily biasing predictions toward the dominant FCC+BCC (blue) region, whereas the physics-informed model consistently converges toward the true decision boundaries across all phase regions.

\begin{figure*}
    \centering
    \includegraphics[width=0.85\linewidth]{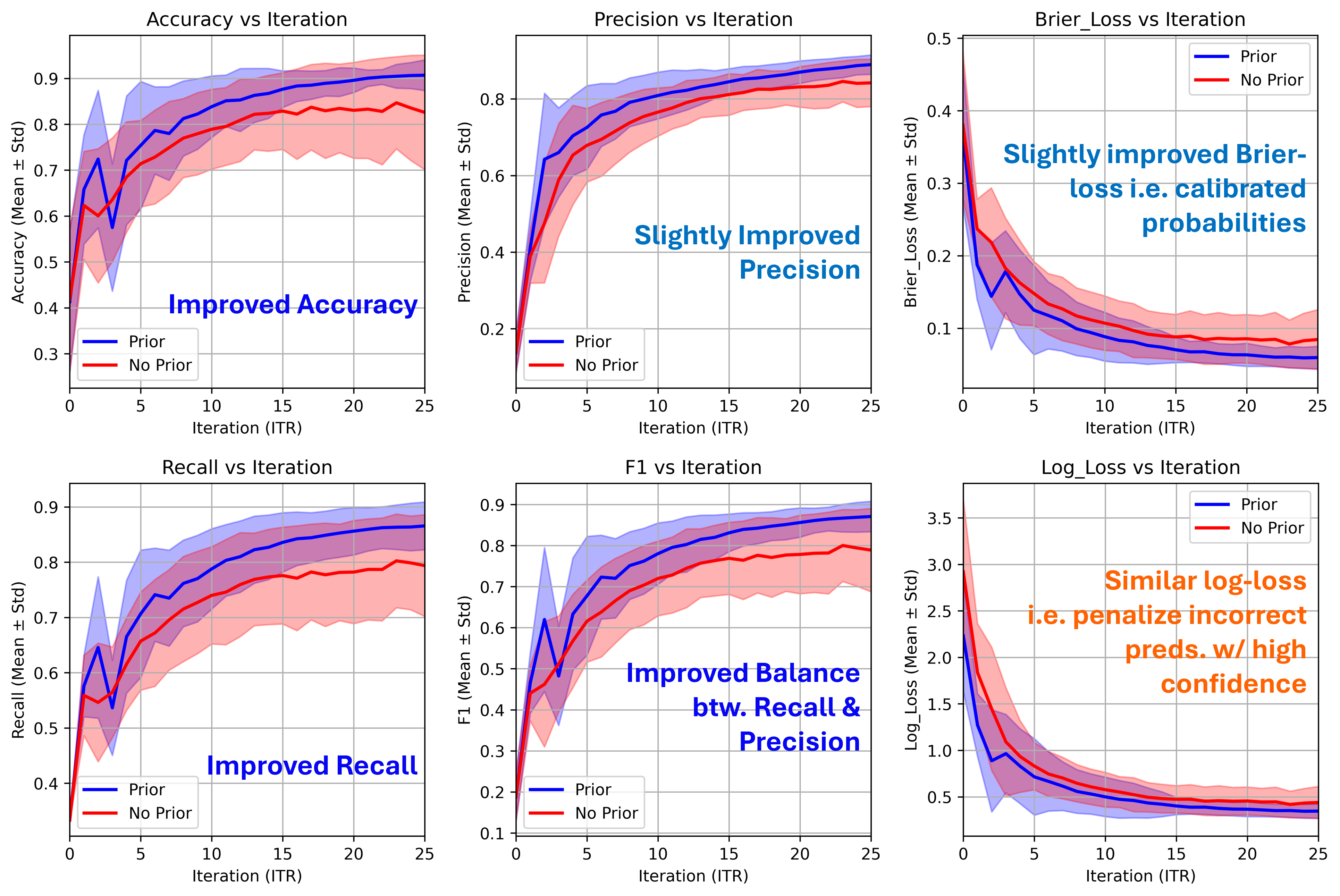}
    \caption{Active learning performance metrics averaged over 200 campaigns. The plotted results show the average error metrics with their standard deviations, providing a more reliable assessment of AL method performance and progression.}
    \label{fig:active_learning_results_stats}
\end{figure*}

Running a single AL campaign is insufficient for benchmarking because a favorable or unfavorable random initialization could unduly influence the results. To address this, we report the distribution of metrics across multiple AL campaigns as a function of iteration, providing a more robust assessment of each method's average performance and progression. Specifically, we run 200 AL campaigns, each with a budget of 25 queries of the ground truth. For each campaign, the six classification metrics described in Section \ref{sec:error_metrics} were recorded at each iteration. The average error metrics and their standard deviations, as a function of AL iteration, are plotted in Figure \ref{fig:active_learning_results_stats}.

The proposed method (blue) shows improved accuracy on average, indicating better overall performance compared to the control model. Furthermore, the standard deviation of accuracy decreases in later iterations, suggesting that the method consistently achieves higher accuracy and is robust to random initializations. In contrast, the control model (red) exhibits a wide accuracy standard deviation that even increases slightly in later iterations, indicating that its performance is less consistent over time and more sensitive to the initial `seed query' of the AL scheme.

\section{Case Study: Informative Priors for Continuous Constraint Satisfaction}

In alloy design, the goal is often to identify an alloy that meets all specified thresholds with high confidence rather than to find the absolute optimal alloy in terms of individual properties \cite{ABUODEH201841}. We contend that, in most real-world examples, quantifying the probability of meeting critical constraints \cite{ghoreishi2019multi} is more important than maximizing any single property. Using the methods described in Section \ref{sec:cont_data}, we develop an active learning scheme to identify the set of W-Nb-Ta alloys with a yield strength exceeding 100 MPa at $1300^\circ \text{C}$ using as few queries as possible. This constraint is adapted from the performance requirements of ARPA-e's ULTIMATE program, which strongly motivates this case study \cite{arroyave2024ultimate}. The confidence that an alloy meets this threshold is represented by the probability mass of the predicted normal distribution (from the GPR model) that falls below 100 MPa. Figure \ref{fig:cont_class_demo} provides a visual demonstration of this classification. Although this is a synthetic problem, it is motivated by previous works \cite{VELA2023118784,VELA2023119351}. Our study demonstrates that equipping GPR models with physics-informed prior mean functions accelerates the identification of alloys that satisfy the yield strength threshold.

\subsection{Ground-Truth and Prior Models for High-Temperature Yield Strength}\label{sec:cont_al_priors_truths}

The Curtin-Maresca model provides a mechanistic framework for predicting the yield strength of BCC high-entropy alloys (HEAs) \cite{maresca2020mechanistic}. Rooted in dislocation theory, this model accounts for the influence of atomic-scale heterogeneities inherent in multicomponent alloys. Specifically, the yield strength is attributed to the resistance encountered by dislocations as they move through a heterogeneous lattice. Such lattice heterogeneities arise from variations in atomic size, elastic modulus, and other local properties due to the random distribution of constituent elements in the alloy.

The critical resolved shear stress is calculated based on the statistical interactions between dislocations and local obstacles, incorporating both temperature and strain rate dependencies. The model employs the following equation to estimate the yield stress:

\begin{equation}
\tau_\text{y}(T, \dot{\varepsilon}) = \tau_{\text{y}0} \exp \left[-\frac{1}{0.55}\left(\frac{k T}{\Delta E_b} \ln \frac{\dot{\varepsilon}_0}{\dot{\varepsilon}}\right)^{0.91}\right],
\label{eq:Maresca-Curtin_Approx_a}
\end{equation}

where \(k\) is the Boltzmann constant, \(T\) is the absolute temperature, \(\tau_{\text{y}0}\) is the zero-temperature shear stress, and \(\Delta E_b\) is the energy barrier for the motion of individual dislocation segments. The strain rate \(\dot{\varepsilon}\) is the applied value, typically set to \(\dot{\varepsilon} = 10^{-3} \, \text{s}^{-1}\), and \(\dot{\varepsilon}_0\) is the reference strain rate, estimated to be \(\dot{\varepsilon}_0 = 10^{4} \, \text{s}^{-1}\). This equation provides a lower-bound estimate for tensile yield strength, as validated in recent studies \cite{baruffi2022screw, VELA2023119351}.

In this work, the Maresca-Curtin model queried at 1300$^{\circ}$C was used as the ground truth for high-temperature yield strength. The model queried at 25$^{\circ}$C was considered the prior. While this is only a toy problem, it emulates a scenario where room-temperature yield strength serves as a proxy for high-temperature yield strength. This prior is updated iteratively.

\subsection{Gaussian Processes and Active Learning Parameters}

The active learning framework for continuous properties used a Gaussian Process for the surrogate model and maximum Shannon entropy for the aquisition function \cite{chen2020active}. The GPRs used in the framework employed the RBF kernel. Since this case involved a straightforward binary classification problem using an analytical model as the ground truth, the White Noise (WN) kernel was omitted. In this study, the RBF kernel models how the 1300$^{\circ}$C yield strength varies with composition in the Nb-Ta-W alloy system. As in previous case studies, the kernel hyperparameters were optimized by maximizing the log-marginal likelihood using the L-BFGS-B algorithm in scikit-learn. To ensure robust optimization, each GPR underwent 50 optimizer restarts.

For the RBF kernel, the optimization was restricted to length scales between 5 atomic percent (at\%) and 100 at\%. This range was chosen based on the Nb-Ta-W alloy space’s sampling resolution of 5 at\% and the observation that barycentric spaces typically do not exhibit length scales beyond 100 at\%. These constraints ensured that the kernel parameters remained physically meaningful and aligned with the characteristics of the sampled data.

\subsection{Continous Active Learning Case Study Results}

To demonstrate the effect of a physics-informed prior during active learning (AL) of constraint boundaries for continuous properties, we equipped one AL scheme with a prior mean function and benchmarked it against an AL scheme without a prior mean function. As mentioned in Section \ref{sec:cont_al_priors_truths}, the ground truth is provided by the Maresca-Curtin model queried at 1300$^{\circ}$C, representing a difficult-to-attain high-temperature tensile measurement. In contrast, the prior in this work is obtained by querying the Maresca-Curtin model at room temperature, representing an easier-to-obtain value.

The model with a physics-informed prior outperforms the model without a prior during the initial iterations of the AL campaign. For example, in iteration 1, the yield strength prediction from the vanilla GPR is constant across the design space, meaning that all alloys receive the same prediction. However, the GPR with the informative prior exhibits a more complex prediction even when provided with only a single data point. The initial predicted decision boundary (i.e., the threshold for alloys having high-temperature yield strength greater than the target value) is more accurately defined. An example of this is shown in Figure \ref{fig:cont_al_terns}.
Both AL schemes were initialized 200 times and ran for 15 iterations. The average performance metrics for each model were plotted as a function of iteration and are shown in Figure 6. 

\begin{figure*}[]
    \centering
    \includegraphics[width=0.85\linewidth]{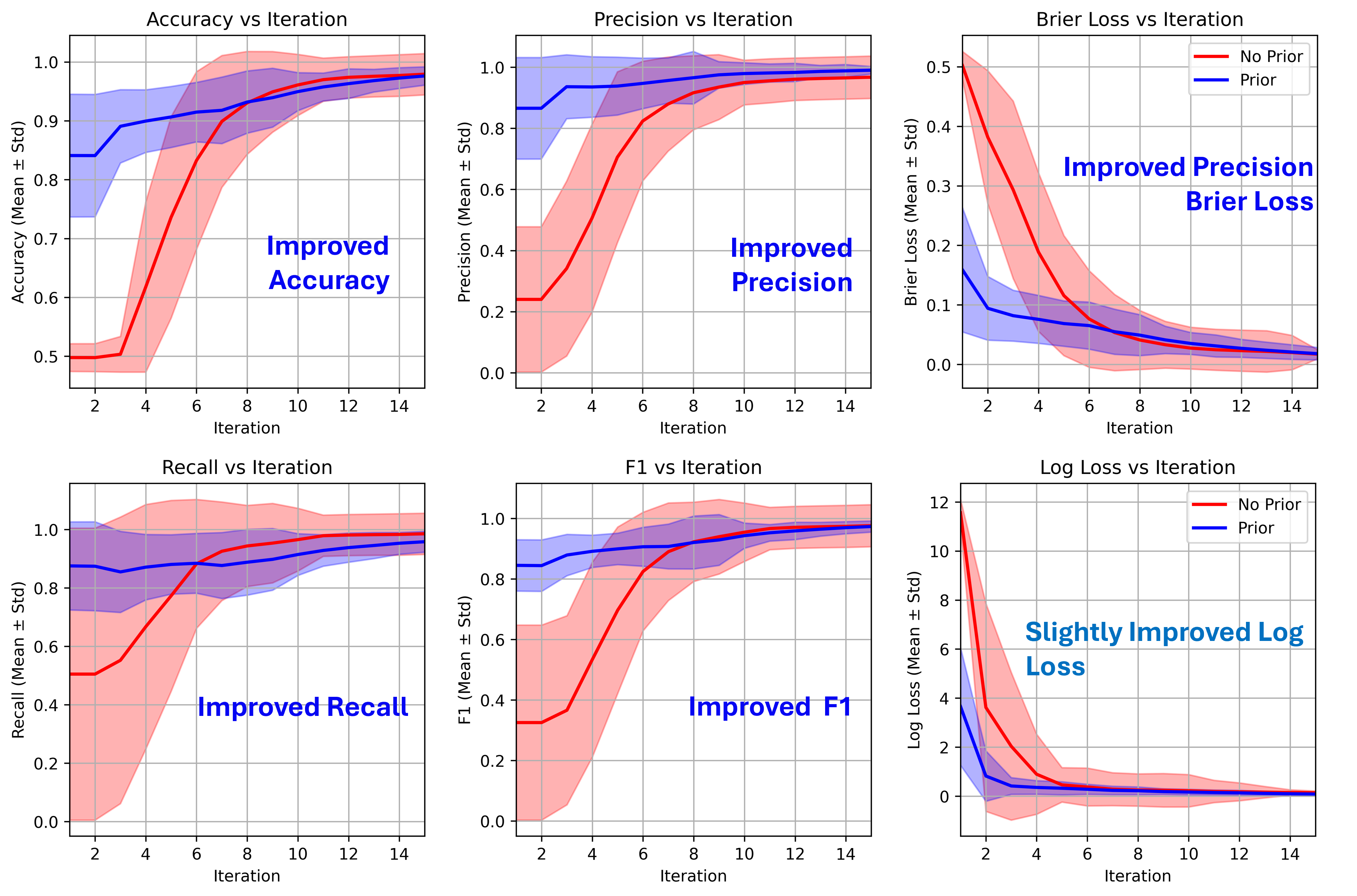}
    \caption{Average performance metrics for continuous property classification (set of W-Nb-Ta alloys where the yield strength exceeds 100 MPa at $1300^\circ \text{C}$) using a GPR. The average metrics for 200 campaigns are plotted as a function of iteration. The blue line represents the average metrics for the model with a physics-informed prior, while the red line represents the average metrics for the model without a prior. The shaded regions show one standard deviation above and below the mean. }
    \label{fig:continuous_classification_metrics}
\end{figure*}

\begin{figure*}[]
    \centering
    \includegraphics[width=1\linewidth]{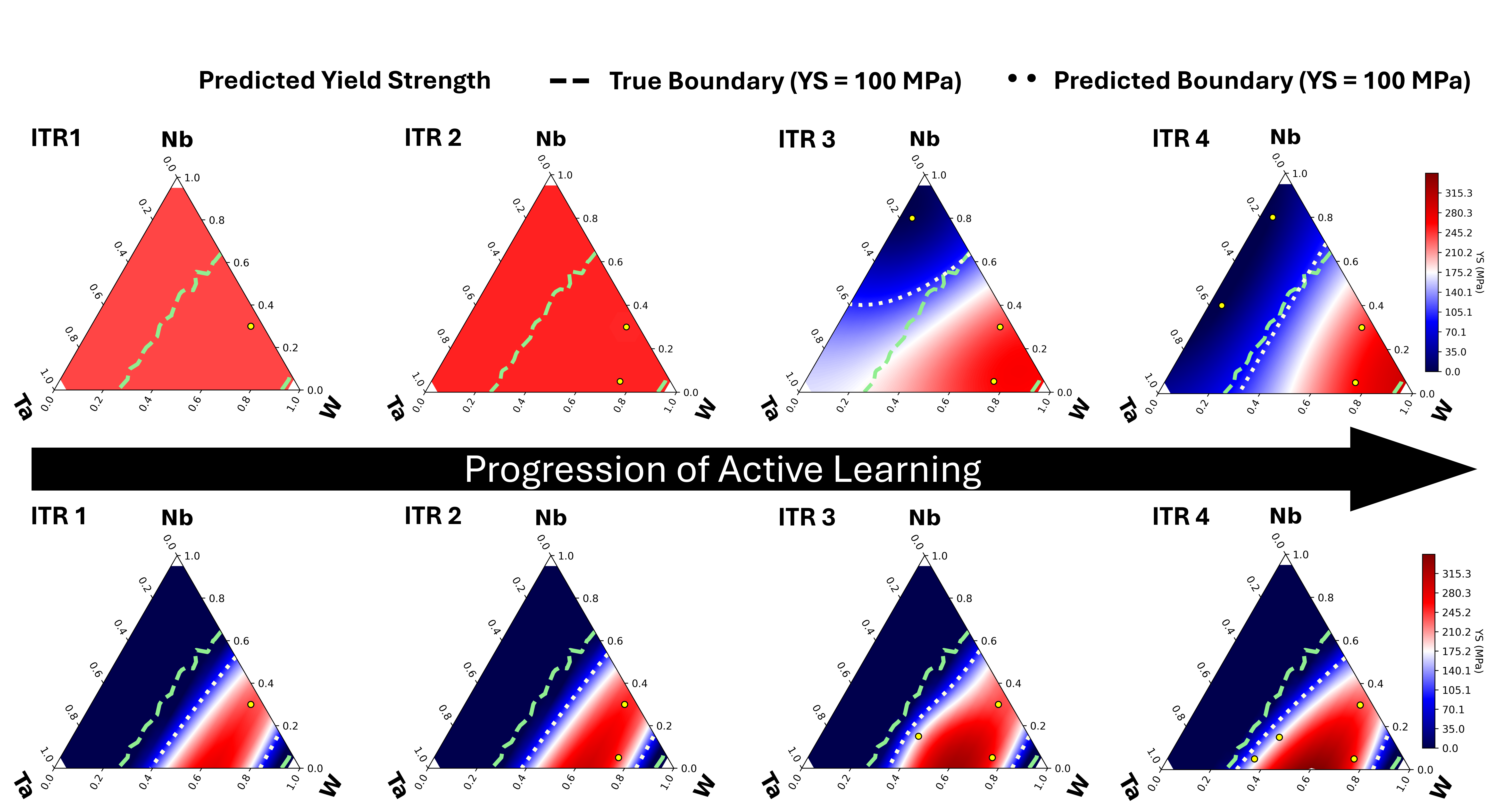}
    \caption{Comparison of vanilla and physics-informed Bayesian active learning for yield strength predictions for W-Nb-Ta alloys at 1300$^{\circ}$C. The top row displays the vanilla AL scheme, and the bottom row illustrates the physics-informed AL scheme. The dashed line represents the true boundary where the yield strength is 100 MPa, while the dotted line indicates the predicted boundary based on the active learning scheme. The physics-informed AL scheme outperforms the vanilla AL scheme for the initial iterations.}
    \label{fig:cont_al_terns}
\end{figure*}

For the first seven iterations, the model with prior data exhibits higher average accuracy and recall. In addition, its average Brier loss and average $F_1$ score are higher for the first eight iterations. The average precision is consistently higher, and the average log loss is consistently lower for the model with prior data. Although the confidence intervals for recall overlap between the two models, the model without prior knowledge shows a notably high standard deviation in recall—exceeding its mean recall value in the first iteration. For all other metrics, the confidence intervals of the two models do not overlap during the first two to four iterations, and the standard deviation is initially lower for the model with a prior.

\section{Conclusion}

In materials design, objectives and constraints play distinct yet complementary roles. Objectives represent desirable material properties that we seek to optimize, while constraints define non-negotiable requirements that a material must satisfy, typically ensuring that it meets minimum performance standards. In previous work \cite{VELA2023119351,MORCOS2024104545}, we demonstrated that Bayesian optimization for property optimization (i.e., maximizing or minimizing objective properties) can be accelerated by incorporating informative priors. In this study, we extend this concept to classification and the active learning of decision boundaries. Specifically, we enhanced Gaussian Process Classifiers (GPCs) with physics-informed priors to make the exploration of material design spaces both more efficient and cost-effective. Our case studies demonstrate that physics-informed prior mean functions can improve the predictive performance of GPCs in alloy design. 

The novelty of our work is twofold. First, for categorical data, this is the first instance in materials science—or, to the best of our knowledge, any related field—where informative prior mean functions have been incorporated into GP classifiers for categorical variables. Second, for continuous data, while GP regressors have been used to classify whether certain properties exceed specific thresholds and, separately, enhanced with informative priors to improve regression, this is the first time both approaches have been combined. Our physics-informed GP regressor predicts the probability that a material meets a given threshold, enabling physics-informed probabilistic predictions within constraint-satisfaction and optimization schemes.

Given the improvements in active learning-based discovery demonstrated in our case studies, we conclude that incorporating physics-informed priors into the alloy design workflow has the potential to significantly reduce computational and experimental costs while enhancing model accuracy and efficiency. The proposed methodology aligns with recent initiatives focused on Integrated Computational Materials Engineering (ICME)-enabled closed-loop design platforms and autonomous materials discovery. Moreover, the approach is easily implemented using only Scikit-Learn and open-access code, ensuring broad accessibility.





\section*{Code Availiability}

The code associated with this work is available at the following repository: \href{https://doi.org/10.24433/CO.5262732.v1}{DOI: 10.24433/CO.5262732.v1}.

\section*{Declaration of Generative AI and AI-assisted technologies in the writing process}

During the preparation of this work the authors used GPT-4-turbo in order to ideate/brainstorm alternative sentence structures and stylistic choices in limited sections of the paper. After using this tool/service, the authors reviewed and edited the content as needed and take full responsibility for the content of the publication.

\section*{Declaration of Competing Interest}
The authors declare that they have no known competing financial interests or personal relationships that could have appeared to influence the work reported in this paper.

\section*{Acknowledgements}
We acknowledge the support from the U.S. Department of Energy (DOE) ARPA-E ULTIMATE Program through Project DE-AR0001427. RA also acknowledges the Army Research Laboratory (ARL) for support through Cooperative Agreement Number W911NF-22-2-0106, as part of the High-throughput Materials Discovery for Extreme Conditions (HTMDEC) program as supported by the BIRDSHOT Center at Texas A\&M University. BV acknowledges the support of NSF through Grant No. 1746932 (GRFP) and 1545403 (NRT-D3EM). Computations were conducted at the Texas A\&M University High-Performance Research Computing (HPRC) facility.

\section*{CRediT authorship contribution statement}
\textbf{Christofer Hardcastle}: Writing – original draft, Writing – review \& editing, Visualization, Software, Investigation, Formal analysis, Data curation. \textbf{Ryan O'Mullan}: Writing – original draft, Visualization, Software, Validation, Investigation, Formal analysis, Data curation. \textbf{Raymundo Arróyave}: Writing – review \& editing, Project administration, Funding acquisition. \textbf{Brent Vela}: Writing – original draft, Writing – review \& editing, Visualization, Software, Investigation, Formal analysis, Data curation,
Conceptualization, Methodology, Supervision.





\end{document}


\title{Supplemental Material: Physics-Informed Gaussian Process Classification for Constraint-Aware Alloy Design}

\begin{frontmatter}

\author[1]{Christofer Hardcastle}
\author[1]{Ryan O'Mullan}
\author[1,2,3]{Raymundo Arr\'oyave}
\author[1]{Brent Vela \corref{cor1}}

\address[1]{Department of Materials Science and Engineering, Texas A\&M University, College Station, TX 77843, USA}
\address[2]{J. Mike Walker '66 Department of Mechanical Engineering, Texas A\&M University, College Station, TX 77843, USA}
\address[3]{Wm Michael Barnes '64 Department of Industrial and Systems Engineering, Texas A\&M University, College Station, TX 77843, USA}

\end{frontmatter}

\begin{figure}[htb!]
    \centering
    \includegraphics[width=0.5\linewidth]{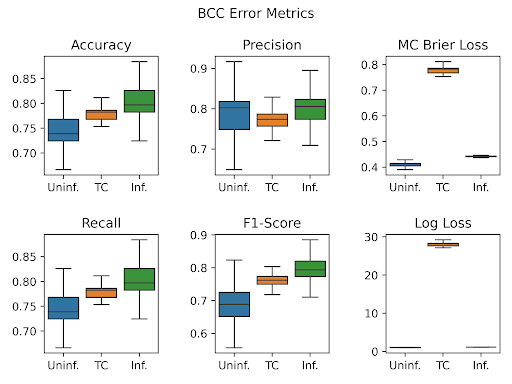}
    \caption{Model errors for the standard GPC (Uninf.), Thermo-calc (TC), and the GPC with the physics-informed prior (Inf.) when predicting for BCC alloys.}
    \label{fig:dist_BCC}
\end{figure}

\begin{figure}[htb!]
    \centering
    \includegraphics[width=0.5\linewidth]{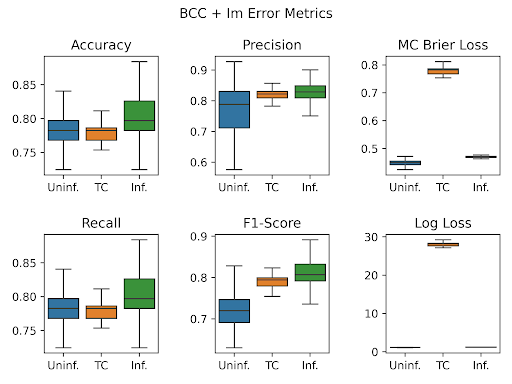}
    \caption{Model errors for the standard GPC (Uninf.), Thermo-calc (TC), and the GPC with the physics-informed prior (Inf.) when predicting for BCC+Sec. alloys.}
    \label{fig:dist_BCCsec}
\end{figure}

\begin{figure}[htb!]
    \centering
    \includegraphics[width=0.5\linewidth]{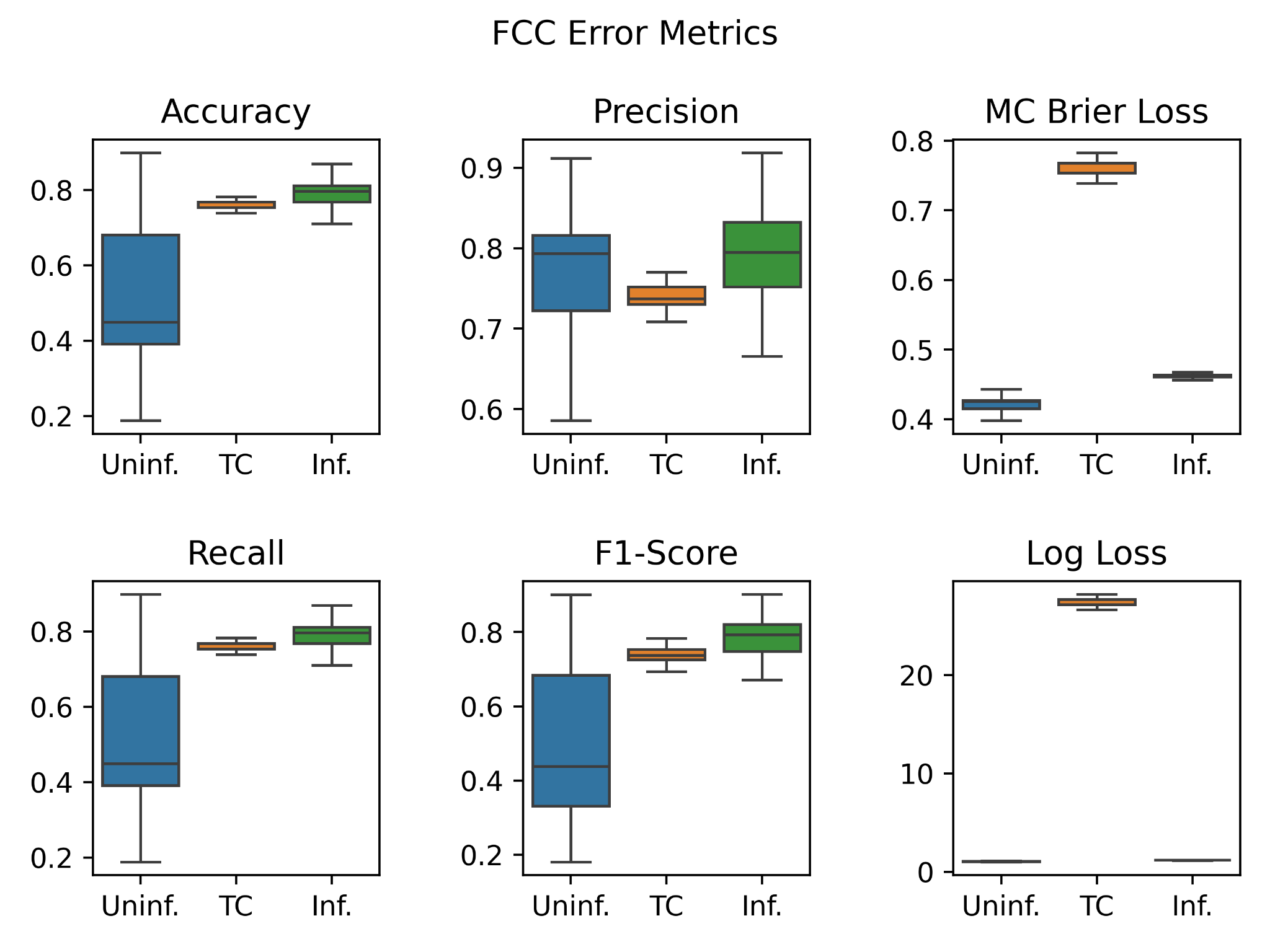}
    \caption{Model errors for the standard GPC (Uninf.), Thermo-calc (TC), and the GPC with the physics-informed prior (Inf.) when predicting for FCC alloys.}
    \label{fig:dist_FCC}
\end{figure}

\begin{figure}[htb!]
    \centering
    \includegraphics[width=0.5\linewidth]{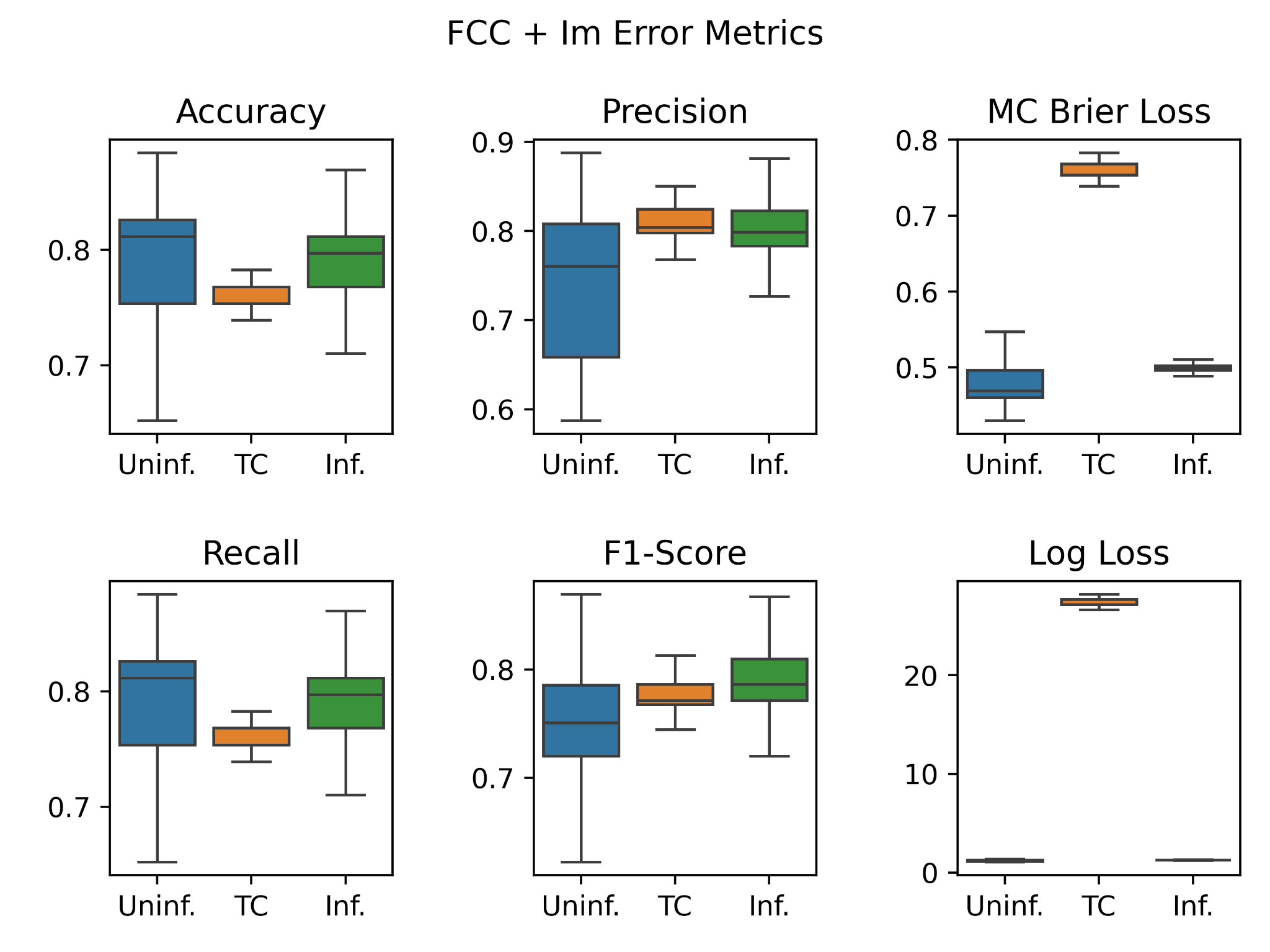}
    \caption{Model errors for the standard GPC (Uninf.), Thermo-calc (TC), and the GPC with the physics-informed prior (Inf.) when predicting for FCC+Sec. alloys.}
    \label{fig:dist_FCCsec}
\end{figure}

In Section 3.4 of the main text, we benchmarked the predictive ability of the proposed method against two control methods. Specifically, the proposed approach was a physics-informed GPC, while the control methods were: (1) a GPC with a constant prior mean function and (2) Thermo-Calc CALPHAD predictions using the TCHEA6 thermodynamic database \cite{tchea6}. For benchmarking, we employed stratified Monte Carlo cross-validation, generating 500 random train/test splits with an 80\%/20\% ratio. To evaluate overall predictive performance across all classes in the multi-class setting, we presented box-and-whisker plots of each error metric based on these cross-validation splits, considering the four-class case. In this Supplemental Material, we extend this analysis to one-vs-rest classifiers, reporting error metric distributions for distinguishing each individual class from all others. Figures \ref{fig:dist_BCC}-\ref{fig:dist_FCCsec} summarize the results for the following one-vs-rest cases:
\begin{itemize}
    \item Single Phase BCC (BCC)
    \item BCC with Secondary Phases (BCC+Sec.)
    \item Single Phase FCC (FCC)
    \item FCC with Secondary Phases (FCC+Sec.)
\end{itemize}

When predicting single-phase BCC and BCC with secondary phases (Figures \ref{fig:dist_BCC} and \ref{fig:dist_BCCsec}), the model with a physics-informed prior outperformed both the standard model and TC in precision, recall, and F1-score, with a slight improvement in precision. However, log loss scores were similar between the physics-informed and standard models, while the physics-informed model performed slightly worse in terms of Brier loss. Overall, the proposed method outperformed the control models in 4 out of 6 error metrics, with its only shortcomings in the probabilistic error metrics, where vanilla GPCs performed slightly better. When considering all error metrics simultaneously, it is evident that the physics-informed GPCs perform the best.

For FCC alloys (Figure \ref{fig:dist_FCC}, the median accuracy, recall, and F1 of the physics-informed GPCs is comparable to that of Thermo-Calc. However, Thermo-Calc exhibits a narrower accuracy, recall, and F1 distributions, indicating greater consistency in distinguishing between single-phase FCC and non-FCC phases. The percision distribution of the physics-informed GPCs outperforms that of the Thermo-Calc. The physics-uninformed GPC has the worst performance with regard to classification these deterministic classification metrics. Interestingly, Thermo-Calc excels with regards to deterministic classification metrics, it performs poorly with regards to probabalistic error metrics whereas the GPCs (both informed and uninformed) both perform well. With all metrics considered, the physics-informed GPCs perform the best.
 
For FCC plus secondary phase predictions (Figure \ref{fig:dist_FCCsec}), the median accuracy, recall, Brier loss and Log loss values of the informed GPC were comparable (if not slightly lower) than the uninformed GPC. However, the variance in these error distribution was lower in the case of the informed GPC, indicating more consistent performance. That is to say, in some cases the uninformed model will perform exceptionally well, and in some cases it will perform exceptionally poorly.

To summarize, while the proposed method does not outperform the control methods with regard to every error metric in every one-vs-rest classification senario, when the error metrics are considered holistically, it is evident that the physics-informed GPC outperform uninformed GPCs and CALPHAD predictions.

\FloatBarrier
